\documentstyle[aps,epsfig]{revtex}
\begin{document}

\title{ Alpha decay half-lives of new superheavy elements }

\author{P. Roy Chowdhury$^1$\thanks{E-mail:partha.roychowdhury@saha.ac.in}, C. Samanta$^{1,2}$}
\address{ $^1$ Saha Institute of Nuclear Physics, 1/AF Bidhan Nagar, Kolkata 700 064, India }
\address{ $^2$ Physics Department, Virginia Commonwealth University, Richmond, VA 23284-2000, U.S.A. }

\author{D.N. Basu}
\address{Variable  Energy  Cyclotron  Centre,  1/AF Bidhan Nagar,
Kolkata 700 064, India}
\date{\today }
\maketitle
\begin{abstract}

      The lifetimes of $\alpha$ decays of the recently produced isotopes of the elements 112, 114, 116 and the element $^{294}118$ and of some decay products have been calculated theoretically within the WKB approximation using microscopic $\alpha$-nucleus interaction potentials. These nuclear potentials have been obtained by folding the densities of the $\alpha$ and the daughter nuclei with the M3Y effective interaction, supplemented by a zero-range pseudo-potential for exchange along with the density dependence. Spherical charge distributions have been used for calculating the Coulomb interaction potentials. These calculations provide reasonable estimates for the observed $\alpha$ decay  lifetimes and thus provide reliable predictions for other superheavies.  

\vskip 0.2cm
\noindent
PACS numbers:27.90.+b, 23.60.+e, 21.30.Fe, 21.65.+f, 25.55.Ci 

\end{abstract}

\pacs{ PACS numbers:27.90.+b, 23.60.+e, 21.30.Fe, 21.65.+f, 25.55.Ci }


\section{Introduction}

      The main features which determine the fusion process for the production of superheavy elements (SHE) are the fusion barrier, and related beam energy and excitation energy, the ratio of surface energy versus Coulomb repulsion which determines the fusion probability and which strongly depends on the degree of asymmetry or the reactants (the product $Z_1Z_2$ at fixed $Z_1+Z_2$), the impact parameter and related angular momentum, and the ratio of neutron evaporation versus fission probability of the compound nucleus. In fusion of heavy elements the product $Z_1Z_2$ reaches extremely large values and the fission barrier extremely small values. In addition, the fission barrier is fragile at increasing excitation energy and angular momentum, because it is solely built up from shell effects. For these reasons the fusion of heavy elements is hampered, whereas the fusion of lighter elements is advanced through the contracting effect of surface tension. Recently isotopes of the elements 112, 114, 116 and the element $^{294}118$ have been produced in the fusion-evaporation reactions keeping low excitation energies by irradiations of the $^{233,238}U$, $^{242}Pu$, $^{248}Cm$ \cite{Og05} and $^{249}Cf$ targets \cite{Og03} with $^{48}Ca$ beam at various energies. The observed decays reveal that the dominant decay mode is the $\alpha$ emission. The $\alpha$ decay energies and half-lives of fourteen new $\alpha$ decaying nuclei have been measured. Incidentally, questions have been raised \cite{Ar00} about some of the superheavy element findings \cite{Og99}. In fact, in similar sophisticated experiments at other places \cite{Lo02}, \cite{Gr05} the $\alpha$ cascades were not observed. While one awaits for further experimental verification of such an important discovery, theoretical predictions already existed for such superheavy elements \cite{Ro02} along with their $\alpha$ decay lifetime predictions \cite{Ro04}.   

      In this work, the half lives of new superheavy elements have been determined with microscopic potentials and compared with the existing theoretical and experimental results to test the extent of validity of this formalism. In view of the excellent agreement of this work with the available experimental data, half lives of about eighty new SHE have been predicted. In this framework, the nuclear potentials have been obtained by double folding the $\alpha$ and daughter nuclei density distributions with a density dependent effective interaction. This nuclear interaction energy for the $\alpha$-nucleus interaction has therefore been obtained microscopically. A double folding potential obtained using M3Y \cite{Be77} effective interaction supplemented by a zero-range potential for the single-nucleon exchange is more appropriate because of its microscopic nature \cite{Sa97}. A potential energy surface is inherently embedded in this description. The semirealistic explicit density dependence \cite{Ko84} into the M3Y effective interaction has been employed to incorporate the higher order exchange and Pauli blocking effects. The penetrability of the pre-scission part of the potential barrier provides the $\alpha$ cluster preformation probability \cite{Po91}. Theoretical calculations in terms of quantum mechanical barrier penetrability using microscopically obtained nuclear potentials have been provided in the present work. Observed lifetimes of the fourteen $\alpha$ decays originating from the isotopes of the synthesized new elements 112, 114, 116 are in reasonable agreement with the theoretical estimates. Recent theoretical predictions \cite{Zh05} for the lifetimes of the $\alpha$ decay chains of superheavy element 115 also agree with the present calculations \cite{BA04} which provided consistent estimates for the observed lifetimes \cite{Og04} of the consecutive $\alpha$ decay chains of the superheavy element 115.  

      Based on the present calculations which  provide reasonable estimates for the observed $\alpha$ decay  lifetimes of many newly synthesized elements and therefore expected to be effective predictors of the half-lives in the region of the heaviest elements, values from years to microseconds have been calculated for various isotopes. This wide range of half-lives encourages the application of a wide variety of experimental methods in the investigations of SHE's from investigation of chemical properties of SHE's using long-lived isotopes, to the atomic physics experiments on trapped ions and to the safe identification of short lived isotopes by recoil separation techniques. 

\section{ The density dependent effective interaction }

      The M3Y interaction has been derived by fitting its matrix elements in an oscillator basis to those elements of the G-matrix \cite{Sa79} obtained with the Reid-Elliott soft-core nucleon-nucleon (NN) interaction. The ranges of the M3Y forces were chosen to ensure a long-range tail of the one-pion exchange potential as well as a short range repulsive part simulating the exchange of heavier mesons. The zero-range potential represents the single-nucleon exchange term while the density dependence accounts for the higher order exchange effects and the Pauli blocking effects. The general expression for the density dependent M3Y effective interaction supplemented by a zero-range potential for the single-nucleon exchange (DDM3Y) is given by 

\begin{equation}
 v(s,\rho, E) = t^{M3Y}(s, E) g(\rho, E) = C t^{M3Y} (1 - \beta(E)\rho^{2/3}) 
\label{seqn1}
\end{equation}   
\noindent
where $\rho$ is the nucleonic density and the M3Y effective interaction potential supplemented by a zero-range potential $t^{M3Y}$ is given by \cite{Ko84} 

\begin{equation}
  t^{\rm M3Y} = 7999 \frac{e^{ - 4s}}{4s} - 2134\frac{e^{- 2.5s}}{2.5s} + J_{00}(E) \delta(s)
\label{seqn2}
\end{equation}   
\noindent
where the zero-range potential $J_{00}(E)$ representing the single-nucleon exchange is given by 

\begin{equation}
 J_{00}(E) = -276 (1 - 0.005 E/A ) (MeV.fm^3)
\label{seqn3}
\end{equation}   
\noindent 

This density dependent M3Y effective NN interaction supplemented by the zero-range potential is used to determine the nuclear matter equation of state. The equilibrium density of the nuclear matter is determined by minimizing the energy per nucleon. The density dependence parameters have been fixed by reproducing the saturation energy per nucleon and the saturation density of spin and isospin symmetric cold infinite nuclear matter. Although the density dependence parameters for single folding can be determined from the nuclear matter calculations and used successfully for proton radioactivity and scattering  \cite{Ba05}, the transition to double folding is not straightforward. The parameter $\beta$ can be related to mean free path in nuclear medium, hence its value should remain same $\sim 1.6 fm^2$ as obtained from nuclear matter calculations \cite{Ba04} while the other constant $C$ which is basically an overall normalisation constant may change. The value of this overall normalisation constant has been kept equal to unity which has been found $\sim 1$ \cite{Ba03} from optimum fit to a large number of alpha decay lifetimes. Since the density dependence of the effective projectile-nucleon interaction has been found to be fairly independent of the projectile, as long as the projectile-nucleus interaction is amenable to a single-folding prescription, implies that in a double folding model, the density dependent effects on the nucleon-nucleon interaction can be factorized into a target term times a projectile term \cite{Sr83}. The general expression for the DDM3Y realistic effective NN interaction to be used to obtain the oft-quoted double-folding nucleus-nucleus interaction potential is given by  

\begin{equation}
  v(s,\rho_1,\rho_2,E) = t^{M3Y}(s,E)g(\rho_1,\rho_2,E)
\label{seqn4}
\end{equation}   
\noindent
where the density dependence term $g(\rho_1, \rho_2, E)$ has now been factorized into a target term times a projectile term \cite{Sr83} as

\begin{equation}
 g(\rho_1, \rho_2, E) = C (1 - \beta(E)\rho_1^{2/3}) (1 - \beta(E)\rho_2^{2/3}).
\label{seqn5}
\end{equation}   
\noindent
The folding model potentials thus obtained by double folding the density distributions $\rho_1$ of the $\alpha$ and $\rho_2$ of the daughter nuclei with such a factorized density dependent M3Y-Reid-Elliott effective interaction, along with a zero-range potential representing the potential arising due to the single-nucleon exchange, have been used successfully to estimate the half lives of the $\alpha$ radioactivity lifetimes of the newly synthesized elements and their isotopes.

\section{The double folded nuclear potentials and the half lives of $\alpha$ radioactivity}

Double folded nuclear interaction potential between the daughter nucleus and the emitted particle is given by \cite{Sa79}

\begin{equation}
 V_N(R) = \int \int \rho_1(\vec{r_1}) \rho_2(\vec{r_2}) v[|\vec{r_2} - \vec{r_1} + \vec{R}|] d^3r_1 d^3r_2 
\label{seqn6}
\end{equation}
\noindent
where $\rho_1$ and $\rho_2$ are the density distribution functions for the two composite nuclear fragments. The density distribution function in case of $\alpha$ particle has the Gaussian form

\begin{equation}
 \rho(r) = 0.4229 exp( - 0.7024 r^2)
\label{seqn7}
\end{equation}                                                                                                                                           \noindent     
whose volume integral is equal to $A_\alpha ( = 4 )$, the mass number of $\alpha$-particle. Since the experimental charge density distributions in case of the heavier nuclei can be well described by the two parameter Fermi function \cite{Fo69} and since the charge which means the proton (p) and the neutron (n) density distributions should have similar forms due to the same strengths of the n-n and p-p nuclear forces, the matter density distribution for the daughter nucleus can be described by the spherically symmetric Fermi function 

\begin{equation}
 \rho(r) = \rho_0 / [ 1 + exp( (r-c) / a ) ]
\label{seqn8}
\end{equation}                                                                                                                                       \noindent     
where the equivalent sharp radius $r_\rho$, the half density radius $c$ and the diffuseness for the leptodermous Fermi density distributions are given by \cite{Sr74}, \cite{Sr83} 
                        
\begin{equation}
c = r_\rho ( 1 - \pi^2 a^2 / 3 r_\rho^2 ),~~r_\rho = 1.13 A_d^{1/3},~~a = 0.54~fm
\label{seqn9}
\end{equation}
\noindent
and the value of the central density $\rho_0$ is fixed by equating the volume integral of the density distribution function to the mass number $A_d$ of the residual daughter nucleus. 

The distance s between any two nucleons, one belonging to the residual daughter nucleus and other belonging to the emitted $\alpha$, is given by $s = |\vec{r_2} - \vec{r_1} + \vec{R}|$ while the interaction potential between these two nucleons $v(s)$ appearing in eqn.(6) is given by the factorised DDM3Y effective interaction described by eqn.(4) and eqn.(5). The total interaction energy $E(R)$ between the $\alpha$ and the residual daughter nucleus is equal to the sum of the nuclear interaction energy, Coulomb interaction energy and the centrifugal barrier. Thus

\begin{equation}
 E(R) = V_N(R) + V_C(R) + \hbar^2 l(l+1) / (2\mu R^2)
\label{seqn10}
\end{equation}   
\noindent
where $\mu = M_e M_d/M$  is the reduced mass, $M_e$, $M_d$ and $M$ are the masses of the emitted particle, the daughter nucleus and the parent nucleus respectively, all measured in the units of $MeV/c^2$. Assuming spherical charge distribution for the residual daughter nucleus and the emitted nucleus as a point particle, the Coulomb interaction potential $V_C(R)$ between them is given by

\begin{eqnarray}
 V_C(R) =&&(\frac{Z_eZ_de^2}{2R_c}).[3-(\frac{R}{R_c})^2]~for~R\leq R_c, \nonumber\\
            = &&\frac{Z_e Z_d e^2}{R}~~otherwise
\label{seqn11}            
\end{eqnarray}   
\noindent
where $Z_e$ and $Z_d$ are the atomic numbers of the emitted-cluster and the daughter nucleus respectively. The touching radial separation $R_c$ between the emitted-cluster and the daughter nucleus is given by $R_c = c_e+c_d$ where $c_e$ and $c_d$ have been obtained using eqn.(9). The energetics allow spontaneous emission of a particle only if the released energy

\begin{equation}
 Q = [ M - ( M_e + M_d ) ] c^2
\label{seqn12}
\end{equation}
\noindent
is a positive quantity.

The half life of a parent nucleus decaying via $\alpha$ emission is calculated using the WKB barrier penetration probability. The assault frequency $\nu$ is obtained from the zero point vibration energy $E_v = (1/2)h\nu$. The decay half life $T$ of the parent nucleus $(A, Z)$  into a $\alpha$ and a daughter $(A_d, Z_d)$  is given by

\begin{equation}
 T = [(h \ln2) / (2 E_v)] [1 + \exp(K)].
\label{seqn13}
\end{equation}
\noindent
The action integral $K$ within the WKB approximation is given by

\begin{equation}
 K = (2/\hbar) \int_{R_a}^{R_b} {[2\mu (E(R) - E_v - Q)]}^{1/2} dR
\label{seqn14}
\end{equation}
\noindent
where $R_a$ and $R_b$ are the two turning points of the WKB action integral determined from the equations 

\begin{equation}
 E(R_a)  = Q + E_v =  E(R_b)
\label{seqn15}
\end{equation}
\noindent
whose solutions provide three turning points. The $\alpha$ particle oscillates between the first and the second turning points and tunnels through the barrier at $R_a$ and $R_b$ representing the second and the third turning points respectively. Since the released energy $Q$ enters in the action integral which goes to the exponential function in eqn.(13) and the zero point vibration energy $E_v$ being proportional to $Q$, the calculations for the lifetimes become very sensitive to the released energies involved in the decay processes.

\section{Calculations and results}

      The two turning points of the action integral given by eqn.(14) have been obtained by solving eqns.(15) using 
the microscopic double folding potential given by eqn.(6) along with the Coulomb potential given by eqn.(11) and the centrifugal barrier. Then the WKB action integral between these two turning points has been evaluated numerically using eqn.(6), eqn.(10) and eqn.(11). The zero point vibration energies used in the present calculations are the same as that described in reference \cite{Po86} immediately after eqn.(4) and experimental $Q$ values have been used. Moreover, the shell effects are implicitly contained in the zero point vibration energy due to its proportionality with the $Q$ value, which is maximum when the daughter nucleus has a magic number of neutrons and protons. Values of the proportionality constants of $E_v$ with $Q$ is the largest for even-even parent and the smallest for the odd-odd one. Other conditions remaining same one may observe that with greater value of $E_v$, lifetime is shortened indicating higher emission rate. Finally the half lives have been calculated using eqn.(13) and tabulated in Tables-I, II.

      The value of the normalization constant C used in the calculations has been kept fixed and equal to unity. All the calculations have been performed with zero angular momentum transfer. The experimentally measured values for the released energy $Q$ have been used in the calculations. In general the E and A appearing in eqn.(3) are the laboratory energy of the projectile in $MeV$ and the projectile mass number respectively. But for a decay process E/A can be shown to be equal to the [energy measured in $MeV$ in the centre of mass of the emitted particle-daughter nucleus system / $(\mu/m)$] where m is the nucleonic mass in $MeV/c^2$ and for the decay process the energy measured in the centre of mass is equal to the released energy $Q$ in $MeV$. Since the released energies involved in the $\alpha$ decay processes are very small compared to the energies involved in high energy $\alpha$ scattering, the zero-range potential $J_{00}(E)$ is also practically independent of energy for the $\alpha$ decay processes and can be taken as $-276 MeV.fm^3$. 

      The results of the present calculations with the DDM3Y for the lifetimes of $\alpha$ decays of recently produced the isotopes of the new elements 112, 114, 116 and the element $^{294}118$ and of some decay products have been presented in Table-I. The quantitative agreement with experimental data is reasonable. The result for $^{294}118$ is almost underestimated possibly because the centrifugal barrier required for the spin-parity conservation could not be taken into account due to non-availability of the spin-parities of the decay chain nuclei. The term $\hbar^2 l(l+1) / (2\mu R^2)$ in eqn.(10) represents the additional centrifugal contribution to the barrier that acts to reduce the tunneling probability if the angular momentum carried by the $\alpha$-particle is non-zero. Hindrance factor which is defined as the ratio of the experimental $T_{1/2}$ to the theoretical $T_{1/2}$ is therefore larger than unity since the decay involving a change in angular momentum can be strongly hindered by the centrifugal barrier. However, as one can see in Table-I that the theoretical Viola-Seaborg systematics with Sobiczewski constants (VSS) \cite{VSS89} largely overestimate the half lives, as many as for eight cases, showing inconsistencies while the present calculations slightly overestimate only for three cases but still provide much better estimates than that estimated by the VSS systematics. For rest of the cases the experimental uncertainties in the $Q$ values associated with the $\alpha$ decays can almost account for the overestimations of theoretical lifetimes if the upper limits for the experimental $Q$ values instead of the mean value be used for the calculations. A very recent theoretical predictions of the generalized liquid drop model (GLDM)  \cite{Ro02}, \cite{Ro04} for these decay lifetimes have also been listed in Table-I and the disagreements of the results with the experimentally observed half lives are primarily due to use of theoretical $Q$ values which do differ from the experimental ones.    

      The theoretical $Q$ values calulated using twentyeight mass excesses from the latest mass table \cite{My96} have also been listed in Table-I for comparison with the experimental ones. It is very obvious from the table that the results for the half lives are quite sensitive to the uncertainties involved in the experimental $Q$ values used in the present calculations. The theoretical $Q$ values differ substantially from the experimental ones for higher $Z,A$ nuclei and they are therefore not used for the calculating the lifetimes. Although the recent theoretical mass table \cite{My96} used for calculating the theoretical $Q$ values provides excellent estimates for normal nuclei, better mass predictions for superheavies are needed for the successful predictions of possible decay modes and their lifetimes.     

      In the Table-II we provide predictions for the alpha decay lifetimes for a large number of superheavy elements \cite{Sm97}, though there exists many more \cite{Ro00}, which are expected to live long enough to be detected after the synthesis in the present day experimental setup. The theoretical $Q$ values have been calculated based on the macroscopic-microscopic (M-M) model \cite{Sm97}. The lifetime values from years to microseconds have been calculated for various isotopes. It is easy to observe that the predictions for the half lives by the present calculations are lower than those by VSS and by the Viola-Seaborg systematics of reference \cite{Sm97}.  

\begin{table}
\caption{Comparison between experimental and calculated $\alpha$-decay half-lives for zero angular momenta transfers, using spherical charge distributions for the Coulomb interaction and the DDM3Y effective interaction. Lower and upper limits of the theoretical half lives corresponding to upper and lower limits of the experimental $Q$ values are also provided. Present theoretical predictions have been compared with those of generalized liquid drop model (GLDM) [7,8] and with VSS [24] predictions. }
\begin{tabular}{ccccccccc}
Parent & Nuclei & Expt.&Assault frequency&Theory& Expt.& DDM3Y &GLDM &VSS   \\ 
&&&(This Work)&Ref.[25] &&(This Work)&&\\ 
$Z$&$A$&$Q(MeV)$&$10^{20} s^{-1}$&$Q(MeV)$&$T_{1/2}$&$T_{1/2}$&$T_{1/2}$&$T_{1/2}$\\ \hline
&&&&&&& \\
  118 &294 &$11.81\pm0.06$&5.968&12.51& $1.8^{+75}_{-1.3} ms$&$0.66^{+0.23}_{-0.18} ms$ &$0.01 ms [8]$&$0.64^{+0.24}_{-0.18} ms$     \\
&&&&&&& \\
  116 &293  &$10.67\pm0.06$&4.680&11.15& $53^{+62}_{-19} ms$&$206^{+90}_{-61} ms$&$18.2 ms [8]$&$1258^{+557}_{-384} ms$      \\  
&&&&&&& \\
  116 &292  &$10.80\pm0.07$&5.458&11.03& $18^{+16}_{-6} ms$&$39^{+20}_{-13} ms$&$6.9 ms [8]$&$49^{+26}_{-16} ms$    \\
&&&&&&& \\
  116&291 &$10.89\pm0.07$&4.777&11.33&$6.3^{+11.6}_{-2.5} ms$& $60.4^{+30.2}_{-20.1} ms$&$7.2 ms [8]$&$336.4^{+173.1}_{-113.4} ms$  \\  
&&&&&&& \\
  116 &290   &$11.00\pm0.08$&5.559&11.34& $15^{+26}_{-6} ms$&$13.4^{+7.7}_{-5.2} ms$&$1.3 ms [8]$&$15.2^{+9.0}_{-5.6} ms$     \\ 
&&&&&&& \\
  114 &289  & $9.96\pm0.06$&4.369&9.08& $2.7^{+1.4}_{-0.7} s$&$3.8^{+1.8}_{-1.2} s$&$51.5 min [8]$&$26.7^{+13.1}_{-8.7} s$    \\ 
&&&&&&& \\
  114 &288  &$10.09\pm0.07$&5.099&9.39& $0.8^{+0.32}_{-0.18} s$&$0.67^{+0.37}_{-0.27} s$&$63 s [8]$&$0.98^{+0.56}_{-0.40} s$   \\ 
&&&&&&& \\
  114 &287 &$10.16\pm0.06$&4.456&9.53& $0.51^{+0.18}_{-0.10} s$&$1.13^{+0.52}_{-0.40} s$&$2.1 min [8]$&$7.24^{+3.43}_{-2.61} s$  \\ 
&&&&&&& \\
  114 &286  &$10.35\pm0.06$&5.230&9.61& $0.16^{+0.07}_{-0.03} s$&$0.14^{+0.06}_{-0.04} s$&$14.5 s [8]$&$0.19^{+0.08}_{-0.06} s$  \\ 
&&&&&&& \\
  112 &285  &$9.29\pm0.06$&4.075&8.80& $34^{+17}_{-9} s$&$75^{+41}_{-26} s$&$83.5 min [8]$&$592^{+323}_{-207} s$     \\ 
&&&&&&& \\
  112 &283 & $9.67\pm0.06$&4.241&9.22& $4.0^{+1.3}_{-0.7} s$&$5.9^{+2.9}_{-2.0} s$ &$3.8 min [8]$&$41.3^{+20.9}_{-13.8} s$   \\ 
&&&&&&& \\
  110 &279  &$9.84\pm0.06$&4.316&9.89& $0.18^{+0.05}_{-0.03} s$&$0.40^{+0.18}_{-0.13} s$&$0.03 s [7]$&$2.92^{+1.4}_{-0.94} s$     \\ 
&&&&&&& \\
  108 &275  &$9.44\pm0.07$&4.141&9.58& $0.15^{+0.27}_{-0.06} s$&$1.09^{+0.73}_{-0.40} s$ &$0.05 s [7]$&$8.98^{+5.49}_{-3.38} s$   \\ 
&&&&&&& \\
  106 &271 &$8.65\pm0.08$&3.794&8.59& $2.4^{+4.3}_{-1.0} min$&$1.0^{+0.8}_{-0.5} min$&$14.8 s [7]$&$8.6^{+7.3}_{-3.9} min$     \\ 
\end{tabular}

\end{table}

\begin{table}
\caption{Comparison between different theoretically predicted $\alpha$-decay half-lives for zero angular momenta transfers using theoretical $Q$ values from the macroscopic-microscopic model. Present calculations using spherical charge distributions for the Coulomb interaction and microscopic nuclear potentials from double folding nuclear densities with DDM3Y effective interaction have been compared with the VSS [24] predictions and with the Viola-Seaborg estimates used in reference [26]. }

\begin{tabular}{cccccccccccc}
Parent & Nuclei  & VSS & DDM3Y& Viola-Seaborg  & M-M model&Parent & Nuclei  & VSS & DDM3Y& Viola-Seaborg  & M-M model\\
&&Ref.[24]&(This Work)&Ref.[26] &Ref.[26]&&&Ref.[24]&(This Work)&Ref.[26] &Ref.[26]\\ 
$Z$&$A$&  $log_{10}T(s)$  & $log_{10}T(s)$  &  $log_{10}T(s)$ &$Q(MeV)$&$Z$&$A$&  $log_{10}T(s)$  & $log_{10}T(s)$  &  $log_{10}T(s)$ &$Q(MeV)$\\ \hline

 104&274&   9.21&   8.75&   9.35&     6.56&104&276&  12.02&  11.55&  12.18&   6.02\\
 104&278&  14.80&  14.31&  15.00&   5.55&104&280&  17.32&  16.80&  17.56&   5.17\\
 104&282&  17.88&  17.34&  18.13&   5.09&104&284&  21.42&  20.87&  21.74&   4.63\\
 104&286&  23.21&  22.65&  23.57&   4.42&104&288&  24.94&  24.36&  25.28&   4.23\\
 104&290&  14.67&  14.01&  14.88&   5.57&104&292&  17.95&  17.28&  18.25&   5.08\\
&&&&&&&&&&&\\
 106&278&   7.92&   7.49&   8.03&   7.02&106&280&  10.50&  10.03&  10.62&   6.48\\
 106&282&  12.58&  12.09&  12.74&   6.09&106&284&  12.75&  12.23&  12.94&   6.06\\
 106&286&  15.61&  15.06&  15.85&   5.58&106&288&  17.26&  16.70&  17.53&   5.33\\
 106&290&  18.45&  17.87&  18.71&   5.16&106&292&  10.60&   9.98&  10.77&   6.46\\
 106&294&  12.86&  12.21&  13.03&   6.04&&&&&&\\
&&&&&&&&&&&\\
 108&282&   7.13&   6.72&   7.17&   7.39&108&284&   8.63&   8.18&   8.73&   7.05\\
 108&286&   8.59&   8.11&   8.69&   7.06&108&288&  11.25&  10.74&  11.40&   6.51\\
 108&290&  12.74&  12.20&  12.92&   6.23&108&292&  13.58&  13.03&  13.73&   6.08\\
 108&294&   7.35&   6.78&   7.43&   7.34&108&296&   8.91&   8.30&   9.02&   6.99\\
&&&&&&&&&&&\\ 
 110&286&   5.38&   5.00&   5.40&   8.02&110&288&   5.38&   4.98&   5.37&   8.02\\
 110&290&   8.08&   7.64&   8.11&   7.36&110&292&   9.15&   8.67&   9.23&   7.12\\
 110&294&   9.67&   9.15&   9.73&   7.01&110&296&   4.96&   4.47&   4.96&   8.13\\
 110&298&   6.08&   5.54&   6.12&   7.84&&&&&&\\
&&&&&&&&&&&\\
 112&288&   2.44&   2.14&   2.35&   9.06&112&290&   3.07&   2.75&   2.98&   8.87\\
 112&292&   5.57&   5.20&   5.56&   8.17&112&294&   6.03&   5.63&   6.02&   8.05\\
 112&296&   6.27&   5.83&   6.26&   7.99&112&298&   2.77&   2.34&   2.70&   8.96\\
 112&300&   3.65&   3.19&   3.59&   8.70&&&&&&\\
&&&&&&&&&&&\\
 114&290&    .02&   -.17&   -.16&  10.08&114&292&   1.52&   1.28&   1.38&   9.57\\
 114&294&   2.84&   2.55&   2.73&   9.15&114&296&   2.91&   2.59&   2.77&   9.13\\
 114&298&   2.98&   2.63&   2.84&   9.11&114&300&    .45&    .12&    .28&   9.93\\
 114&302&   1.03&    .67&    .87&   9.73&&&&&&\\
&&&&&&&&&&&\\
 116&284&  -6.19&  -6.04&  -6.57&  12.96&116&286&  -4.92&  -4.84&  -5.26&  12.34\\
 116&288&  -3.18&  -3.18&  -3.48&  11.56&116&290&  -2.24&  -2.30&  -2.51&  11.17\\
 116&292&  -1.99&  -2.09&  -2.26&  11.07&116&294&  -1.15&  -1.28&  -1.40&  10.74\\
 116&296&   -.99&  -1.15&  -1.25&  10.68&116&298&   -.99&  -1.18&  -1.24&  10.68\\
 116&300&  -1.02&  -1.23&  -1.26&  10.69&116&302&  -2.68&  -2.87&  -2.96&  11.35\\
 116&304&  -2.24&  -2.47&  -2.52&  11.17&&&&&&\\
&&&&&&&&&&&\\
 118&288&  -5.97&  -5.79&  -6.39&  13.11&118&290&  -4.64&  -4.53&  -5.02&  12.46\\
 118&292&  -4.23&  -4.15&  -4.61&  12.27&118&294&  -4.05&  -4.00&  -4.42&  12.19\\
 118&296&  -3.79&  -3.77&  -4.15&  12.07&118&298&  -3.54&  -3.56&  -3.90&  11.96\\
 118&300&  -3.56&  -3.61&  -3.91&  11.97&118&302&  -3.61&  -3.68&  -3.98&  11.99\\
 118&304&  -4.77&  -4.82&  -5.15&  12.52&&&&&&\\
&&&&&&&&&&&\\
 120&292&  -6.40&  -6.14&  -6.88&  13.59&120&294&  -6.07&  -5.85&  -6.55&  13.42\\
 120&296&  -6.03&  -5.84&  -6.51&  13.40&120&298&  -5.95&  -5.79&  -6.43&  13.36\\
 120&300&  -5.42&  -5.31&  -5.87&  13.09&120&302&  -5.38&  -5.29&  -5.83&  13.07\\
 120&304&  -5.48&  -5.41&  -5.93&  13.12&120&306&  -6.28&  -6.21&  -6.76&  13.53\\

\end{tabular}

* All the nuclei listed above are either spherical or have very small deformations [26]. 

\end{table}

\section{Summary and conclusion}

      The half lives for $\alpha$-radioactivity have been analyzed with microscopic nuclear potentials obtained by the double folding procedure using DDM3Y effective interaction. This procedure of obtaining nuclear interaction potentials is based on profound theoretical basis. The results of the present calculations using DDM3Y are in good agreement with the published experimental data for the half lives of the alpha decays from the isotopes of the elements 112, 114, 116, from the element $^{294}118$ and from some decay products. As some of these experimental data await further experimental verification, these theoretical predictions are expected to provide useful guideline. Lifetime estimates from present calculations are lower than those of Viola-Seaborg systematics. The released energies $Q$, to which the calculations are quite sensitive, when calculated from the microscopic-macroscopic model masses \cite{My96} do not provide excellent agreements with those observed for superheavies. Nevertheless, the positive decay $Q$ values \cite{My96} support these $\alpha$ decay modes. Present calculations demonstrate its success of providing reasonable estimates for the lifetimes of nuclear decays by $\alpha$ emissions for the domain of superheavy nuclei.


\end{document}